\documentclass[letterpaper, 10 pt, conference]{ieeeconf}  

\usepackage[utf8]{inputenc}
\usepackage{fourier}
\usepackage{amsmath}
\usepackage[utf8]{inputenc}
\usepackage{amssymb}
\usepackage[utf8]{inputenc}
\usepackage{graphicx}
\usepackage{caption}
\usepackage{amsmath}
\newcommand\norm[1]{\left\lVert#1\right\rVert}
\usepackage{mathtools}
\usepackage[thinc]{esdiff}
\usepackage{xcolor}
\usepackage{epsfig}
\usepackage{subcaption}
\usepackage{hf-tikz}
\usepackage[ruled,vlined]{algorithm2e}
\usepackage{array}
\IEEEoverridecommandlockouts                              

\title{\LARGE \bf
On Linear Time Invariant Systems Analysis via A Single Trajectory: A Linear Programming Approach}

\author{Hassan Abdelraouf$^{1}$, Fahad Albalawi$^{2}$ and Eric Feron$^{3}$
\thanks{*This work was supported by King Abdullah University of Science and Technology (KAUST)}
\thanks{$^{1}$Hassan Abdelraouf is a Ph.D. student with Mechanical Engineering, (KAUST), Thuwal, KSA
        {\tt\small hassan.abdelraouf@kaust.edu.sa}}%
\thanks{$^{2}$ Fahad Albalawi is a research scientist with Electrical and Computer Engineering, KAUST, Thuwal, KSA
        {\tt\small fahad.albalawi@kaust.edu.sa}}%

\thanks{$^{3}$ Eric Feron is with the Faculty of Electrical, Computer, and Mechanical Engineering at KAUST, Thuwal, KSA
        {\tt\small eric.feron@kaust.edu.sa}}%
}

\begin{document}

\maketitle
\thispagestyle{empty}
\pagestyle{empty}

\begin{abstract}
In this note, a novel methodology that can extract a number of analysis results for linear time-invariant systems (LTI) given only a single trajectory of the considered system is proposed. The superiority of the proposed technique relies on the fact that it provides an automatic and formal way to obtain a valuable information about the controlled system by only having access to a single trajectory over a finite period of time (i.e., the system dynamics is assumed to be unknown). At first, we characterize the stability region of LTI systems given only a single trajectory dataset by constructing the associated Lyapunov function of the system. The Lyapunov function is found by formulating and solving a linear programming (LP) problem. Then, we extend the same methodology to a variety of essential analysis results for LTI systems such as deriving bounds on the output energy, deriving bounds on output peak, deriving $\mathbf{L}_2$ and RMS gains. To illustrate the efficacy of the proposed data-driven paradigm, a comparison analysis between the learned LTI system metrics and the true ones is provided.   

\end{abstract}

\section{INTRODUCTION}

Data-driven based control schemes gained a significant attention in the control society over the last decade. In fact, such control paradigms become an attractive alternative to conventional control algorithms \cite{deisenroth2013gaussian,chua2018deep}. This paradigm shift from explicit control techniques to learning-based ones has produced a massive amount of theoretical and practical research works where the majority of these control techniques were obtained from supervised machine learning and reinforcement learning \cite{aswani2013provably,berkenkamp2016safe,beckers2019stable,lederer2020parameter,fisac2018general}. Despite the fact that theoretical certificates for control performance may be derived in various ways, the direct relationship between the generated or collected data and the learning-based control performance is not well-established \cite{lederer2021impact}. Albeit data quality metrics such as entropy have been utilized substantially to guide exploration and control strategies \cite{pukelsheim2006optimal,hennig2012entropy}, they still do not yield direct conclusions into the influence of data on the provable control performance.   

As a remedy, \cite{lederer2020training} utilized Gaussian process priors to generate Lyapunov-based measure that can assess the value of data points with respect to a number of control tasks. The ultimate goal of collecting dataset and assessing their quality for dynamical system is to derive stability and performance certificates that can be used later for different control tasks. As customary, the stability of an equilibrium point of a given dynamic system can be studied through the Lyapunov function. Lyapunov functions can be seen as stability certificates for ordinary differential equations (ODEs) \cite{narendra1994common,mason2007linear}. The problem of finding the Lyapunov function is generally complex, and it has been the scope of many research papers in the control community \cite{ravanbakhsh2017learning}. Analytical techniques are standard tools for constructing Lyapunov functions. Despite the fact that these techniques are mathematically stable and sound, they require significant expertise and manual efforts \cite{abate2020formal}. For LTI systems, which is the focus class of systems for this work, semi-definite programming is sufficient to construct Lyapunov functions due to the fact that Lyapunov functions are inherently quadratic polynomials for LTI systems. However, the knowledge of the underlying dynamical linear systems, i.e., the matrices (A,B,C,D), is always assumed to be known in order to derive the Lyapunov function via semi-definite programming approaches. Such an assumption might not be fulfilled when first principle models cannot be derived
because of the complexity of the actual system. Alternatively, data-driven algorithms can provide an approximate mathematical model from real system measurements.

Motivated by the aforementioned observations, we introduce a formal and automatic methodology for LTI systems analysis where the model of the controlled system is not available and only a single trajectory dataset is provided. Specifically, we formulate the problems of finding the Lyapunov function and observability gramium of LTI systems as a  linear programming (LP) problem given dataset that represents a single system trajectory over a finite period of time. Many problems in control systems analysis like calculating system's RMS gain and output peak are formulated as Semi-definite Programming (SP) and Linear Matrix Inequalities (LMIs) \cite{boyd1994linear} where a full knowledge of the system dynamics is assumed. By using our novel approach, these problems can be solved accurately and efficiently without knowing the system dynamics, and only a single trajectory over a finite period of time is given. Majority of the proposed data-driven based Lyapunov function methods in the literature focus on using Sum of Squares (SOS) methods \cite{papachristodoulou2002construction,papachristodoulou2005tutorial} as well as machine learning techniques like Neural networks. \cite{chang2020neural,chen2021learning}. Nevertheless, all of these methods are model-based and computationally expensive where big data sets are needed for efficient learning. Our approach requires the least data points along the trajectory to learn the Lyapunov function for LTI systems as well as other LTI system analysis metrics. Finally, we test our proposed frameworks by comparing them with the true Lyapunov function as well as the true LTI systems metrics where superior performance of our proposed algorithms is well-established.  
\section{Preliminaries}
\subsection{Class of LTI systems}
We consider a class of LTI systems that can be written in the following state description: 
\begin{align}\label{LTI system}
\begin{split}
    \dot{x} &=A x + B u \\
    z & = Cx + Du 
\end{split}
\end{align}
where $x \in \mathbb{R}^n$ represents the state vector, $u \in \mathbb{R}^m $ is the control input vector, and $z\in \mathbb{R}^p $ is the output state vector. 
\subsection{Notations}

The $2$ norm of a vector $x\in \mathbb{R}^n$ is defined as follows $\left(\sum_{i=1}^{n}\left|x_{i}\right|^{2}\right)^{1 / 2}$. We denote the vector of unique elements of a symmetric matrix $P \in \mathbb{R}^{n \times n}$ as $\text{vec}(P)$ and it is defined as follows:
\begin{equation}\label{vecP def}
    \text{vec}(P) = \begin{bmatrix}
     [P_{11} \quad\dots \quad P_{nn}]^T\\
     [P_{12} \quad \dots \quad P_{1n}]^T\\
     [P_{23} \quad \dots \quad P_{2n}]^T\\
     \vdots \\
     P_{(n-1)n}
     \end{bmatrix} 
\end{equation}

where $\text{vec}(P) \in \mathbb{R}^{n(n+1)/2}$. For instance, when $P \in \mathbb{R}^{2\times2}$ is symmetric, then
$\text{vec}(P)= [ P_{11} \quad P_{22} \quad P_{12}]^T$. In addition, for the two vectors $x,y \in \mathbb{R}^n$ , we define $\oplus$ operator as follows: 
\begin{equation}\label{oplus operator def}
     x \oplus y = \begin{bmatrix}
     [x_1 y_1 \quad\dots \quad x_n y_n]^T\\
     [x_1y_2+x_2y_1 \quad \dots \quad x_1y_n+x_n y_1]^T\\
     [x_2y_3+x_3y_2 \quad \dots \quad x_2y_n+x_n y_2]^T\\
     \vdots \\
     x_{n-1}y_n + x_n y_{n-1}
     \end{bmatrix}
 \end{equation}

\noindent If $z=x \oplus y$, then $z \in \mathbb{R}^{n(n+1)/2}$. For example, if $x,y \in \mathbb{R}^2$, then $x \oplus x = [x_1^2 \quad x_2^2 \quad  2x_1x_2]^T $ and $x \oplus y = [x_1 y_1 \quad x_2 y_2 \quad x_1 y_2 + x_2 y_1]^T $. Using the definitions of both operators $\oplus$ and $\text{vec}(.)$, we can represent the term $x^T P y$ as $(x \oplus y)^T \text{vec}(P)$

Finally, the $\mathbf{L}_2$ norm of a signal $\zeta$ is defined as $\norm{\zeta}_2^2=\int_0^{\infty} \zeta^T \zeta \text{dt} $ and its root-mean square gain $\mathbf{R M S}(\xi)$ is defined as follows: 
\begin{equation}\label{RMS gain def}
\mathbf{R M S}(\xi) \triangleq\left(\limsup _{T \rightarrow \infty} \frac{1}{T} \int_{0}^{T} \xi^{T} \xi d t\right)^{1 / 2}
\end{equation}

\subsection{Outline}
This paper is organized as follows. Section \ref{learning lyapunov } presents using our approach to learn lyapunov function and solve lyapunov equation for LTI system using data along a single trajectory. Sections \ref{learning observability}, \ref{learn bounds on output peak}, \ref{learn L_2 norm}  show how the proposed approach is used to learn bounds on output energy, bounds on output peak and $\mathbf{L}_2$ gain of LTI systems respectively. 
\section{Data-driven Construction for Lyapuonv function for LTI systems} \label{learning lyapunov }
Stability analysis of dynamical systems aims to show that a set of initial states will stay in the neighborhood of an equilibrium point or converge to it. Based on the construction of Lyapunov functions, the stability of equilibrium points, which can form positive invariant sets (i.e., regions of attraction), can be certified. Lyapunov showed that the linear time invariant system is stable (i.e. all trajectories converge to zero asymptotically) if and only if there exists a quadratic function $V(x)=x^T P x$ which is positive $V(x)>0$ and its gradient is negative along the system's trajectories $\dot{V}(x)<0$. These two conditions can be formulated as the following Linear matrix inequalities: $P>0$ and $ P A + A^T P <0$ \cite{boyd1994linear}. Lyapunov inequality in $P$ can be explicitly solved by picking $Q=Q^T >0$ , then solving the linear equation $PA+A^T P =-Q$ for the matrix $ P$ \cite{boyd1994linear}. This method assumes the knowledge of the system dynamic equation of Eq. \ref{LTI system}. For real systems, it may be very difficult to obtain the system dynamics and very expensive to estimate its unknown parameters. Hence, we propose a new method to learn Lyapunov function for LTI systems given only a single trajectory. To the best of our knowledge, our proposed approach is very novel and it has never been introduced in the control society as a tool for LTI system analysis. For stable LTI systems, Lyapunov function is known to be quadratic ($x^T P x$). So, our approach focuses on deriving a $P$ matrix that makes the Lyapunov function positive and its gradient negative at all given data points along the given system trajectory. 

 First, a trajectory for the LTI system is generated  from the initial time $t_0 = 0$ to $T+\text{dt}$ with a time step $\text{dt}$. The states along this trajectory are  $\begin{bmatrix} x(0) &x(1) &\dots &x(N) & x(N+1) \end{bmatrix}$, where $N=T/\text{dt}$ and $x(i)$ represents the state vector at time $i \text{dt}$. Then, the state vector derivative w.r.t. time $\dot{x}(i)$ are calculated numerically using forward finite difference differentiation from $i=0$ to $i=N$. So, the data points $i=0 \dots N$ for state vectors $x$ and state vectors derivative $\dot{x}$ are only used to obtain the unknown parameters of $P$ matrix.  The objective is to find the unknown parameters of $P \in \mathbb{R}^{n \times n}$ which satisfies Lyapunov conditions at every point on the given trajectory:
 \begin{equation}\label{discrete Lyapunov conditions}
 \begin{gathered}
     V(x(i))=x(i)^T P x(i)>0\\
     \frac{d}{\text{dt}} V(x(i))=2x(i)^T P \dot{x}(i)<0
 \end{gathered}
 \end{equation}
for all $i=0,\dots, N$. It is known that $P$ matrix is a symmetric positive definite matrix, so the number of unknown parameters in $P$ is $n(n+1)/2$. Let $p \in \mathbb{R}^{n(n+1)/2}$ is the vector of unknown parameters. So, $p=\text{vec}(P)$ using (\ref{vecP def}). Therefore, the Lyapunov-based conditions of (\ref{discrete Lyapunov conditions}) can be formulated as a set of linear inequalities in  $p$ as: $L_1 p >\epsilon$ and $L_2 p<-\epsilon$ where, 
\begin{equation}\label{L_1}
L_1 = \begin{bmatrix}
(x(0) \oplus x(0))^T \\ 
(x(1) \oplus x(1))^T \\
\vdots \\ 
(x(N) \oplus x(N))^T 
\end{bmatrix}
\end{equation} 
\begin{equation}\label{L_2}
    L_2 = 2\begin{bmatrix}
(x(0) \oplus \dot{x}(0))^T\\ 
(x(1) \oplus \dot{x}(1))^T\\
\vdots \\ 
(x(N) \oplus \dot{x}(N))^T
\end{bmatrix}
\end{equation}
such that $L_1,L_2 \in \mathbb{R}^{(N+1)\times n(n+1)/2}$. The parameter $\epsilon$ is a positive small number. These linear inequalities can be solved trivially by any linear programming solver such as CVX \cite{grant2014cvx} to get $p$ (i.e., the vector of unknown parameters of $P$ matrix that defines our desired Lyapunov function $V(x)$). 

\noindent \underline{\textbf{Numerical example:}}
\noindent We have the following unforced LTI system: $\dot{x}=A x$ where
\begin{equation} \label{system}
    A=\begin{bmatrix} 
    0 & 1\\ -1 &-3 
    \end{bmatrix}
\end{equation}

\begin{figure}[hbt!]
    \centering
    \includegraphics[scale=0.5]{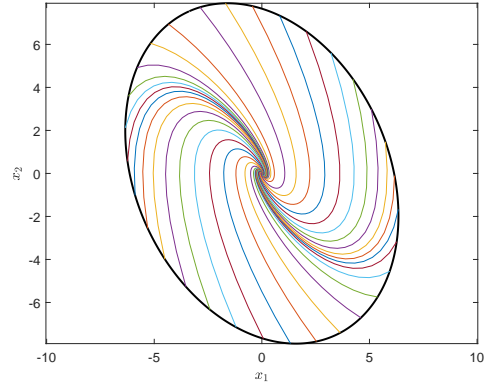}
    \caption{Data-driven Lyapunov function with $V(x)<=1000$}
    \label{fig:1}
\end{figure}

Our approach utilizes only one generated trajectory and then learns the quadratic Lyapunov function $x^T P x$ from data along this trajectory to prove the system stability. The given trajectory starts at $x(0)=[2,2]^T$ to the final time $T=1 \text{sec}$ with time step $dt=0.01 \text{sec}$ and $N=100$. First, the finite difference method is used to get $\dot{x}$ along the given trajectory. Then, the two sets of linear inequalities that represent the satisfaction of Lyapunov conditions at every point on the given trajectory $L_1p>\epsilon$ and $L_2p<-\epsilon$ are solved. Since the dimension of the state vector $x$ is $\mathbb{R}^2$, $L_1$ and $L_2$ can be structured as follows:

\begin{equation}
     L_1 =\begin{bmatrix}
    x_1(0)^2 &x_2(0)^2& 2x_1(0)x_2(0)\\
   \vdots &\vdots&\vdots \\ 
    x_1(N)^2 &x_2(N)^2& 2x_1(N)x_2(N)
    \end{bmatrix}
\end{equation}
\footnotesize{
   \begin{equation}
   L_2 = 2\begin{bmatrix}
    x_1(0)\dot{x_1}(0) & x_2(0) \dot{x_2}(0) & x_1(0)\dot{x_2}(0)+x_2(0)\dot{x_1}(0)\\
    \vdots&\vdots&\vdots\\
    x_1(N)\dot{x_1}(N) & x_2(N) \dot{x_2}(N) & x_1(N)\dot{x_2}(N)+x_2(N)\dot{x_1}(N)
    \end{bmatrix}
   \end{equation}}
   \normalsize
These linear inequalities are solved by Matlab CVX  solver \cite{grant2014cvx} to get the unknown $P$ matrix: 
\begin{equation}
P=
\begin{bmatrix}
  26.4840  & 5.3151\\
   5.3151 & 17.0361
\end{bmatrix}
\end{equation}
Since $P$ is positive definite matrix and $PA+A^TP$ is negative definite, the learned  quadratic function $V(x)=x^T P x$ is a valid Lyapunov function for the LTI system $\dot{x} =Ax $. Such function was derived from dataset along a single trajectory of the considered LTI system. Fig. \ref{fig:1} shows one level set of the learned Lyapunov function ($x^T P x <=1000$) which constitutes a forward invariant set $\Omega_\rho$ where $\rho$ is 1000. Several points at the boundary of the Lyapunov level set are chosen to be initial states for the system trajectories. All the trajectories remain inside the set which demonstrates that the learned function is a true Lyapunov function.

Lyapunov functions for stable LTI systems are not necessarily unique. As a result of such fact, changes on the Lyapunov conditions can be made by the user to improve the numerical stability of the linear program. Lyapunov conditions can be modified to be $L_1p> c_1 l  $ and $L_2 p<-c_2 l$. Where $c_1,c_2 > 0$ and $l \in \mathbb{R}^{(N+1)\times 1}$ is: 

\begin{equation}\label{l}
    l = \begin{bmatrix}
    \norm{x(0)}_2^2\\
    \vdots \\
    \norm{x(N)}_2^2
    \end{bmatrix}
\end{equation}

To improve the robustness of the learned Lyapunov function $V(x)$, the number of data points need to be increased and a probing noise to the given trajectory needs to be augmented.

\subsection{Exact solution of Lyapunov equation from data}

The proposed approach can be employed to solve the Lyapunov equation $PA+A^T P = -Q$ accurately by knowing only the states $x$ and their derivatives w.r.t time evaluated at least at $n(n+1)/2$ points on any given trajectory. where $n$ is the state space dimension. The solution of the Lyapunov equation is based on formulating the problem as an LP problem without the need to know the system dynamics (i.e., $A$ matrix for LTI systems). The matrix $Q \in \mathbb{R}^{n \times n}$ at the right hand side of the Lyapunov function is a user defined symmetric positive definite matrix. Hence, the set of linear inequalities will be: $L_1 p  > \epsilon$, $L_2 p \leq l_Q$ and $L_2 p \geq l_Q$. The vectors $L_1$ and $L_2$ are defined in (\ref{L_1}) and (\ref{L_2}) while $l_Q$ is defined as follows:  
\begin{equation}\label{l_Q}
    l_Q =\begin{bmatrix}
      x(0)^T Q x(0) \\
      x(1)^T Q x(1) \\
      \vdots \\
      x(N)^T Q x(N) 
    \end{bmatrix}
\end{equation}

\noindent \underline{\textbf{Numerical example:}}
From the same state trajectory used in the previous example, we will now pick only three data points to derive the Lyapunov equation of the system of Eq. \ref{system}. The three data points are ($x,\dot{x}$) at the following time instants $0,0.5$ and $1$ sec.  When $Q=-I$, then the sets of linear inequalities in $p$ (i.e., $L_1 p  > \epsilon$, $L_2 p \leq l_Q$ and $L_2 p \geq l_Q$) are as follows: 

\begin{align}
    L_1& = \begin{bmatrix}
        4 &4&8 \\
          4.9627& 0.4626& -3.0304\\
          2.6757&2.1311 & -4.7759  
          \end{bmatrix}   \\
      L_2 &= \begin{bmatrix}
        8 & -32& -24\\
        -3.0304 & 0.2547 &0.0910 \\
         -4.7759& -8.0107& 13.2384
    \end{bmatrix}\\
    l_Q &= \begin{bmatrix}
     8 \\ 
      5.4253\\
      4.8068
    \end{bmatrix}
\end{align}
The solution of these linear inequalities is: 
\begin{equation}
    P = \begin{bmatrix}
    1.8333  & 0.5000 \\ 
     0.5000 & 0.3333
    \end{bmatrix}
\end{equation}
which is exactly the same solution of $PA+A^TP=- I$, yet derived by only using three data points along any given system's trajectory. Such an interesting result shows the efficacy and accuracy of our proposed approach.

The main idea is that $P$ matrix has $n(n+1)/2$ unknown parameters so, we need only $n(n+1)/2$ equations to solve for $P$. These equations represents the satisfaction of Lyapunov second inequality at $n(n+1)/2$ points on the given trajectory. 
\section{Data-driven evaluation for bounds on the output energy} \label{learning observability}
The maximum output energy for LTI system given the initial state is given by: 
\begin{equation}
\max \left\{\int_{0}^{\infty} z^{T} z d t \mid \dot{x}=A x, \quad z=C x\right\}
\end{equation}
Where the initial state $x(0)$ is given. Suppose that there exists a quadratic positive function $V(\zeta)=\zeta^T P \zeta$ such that:
\begin{equation} \label{obser conditions}
 P>0 \text { and } \frac{d}{\text{dt}} V(x) \leq-z^{T} z, \quad \text { for every } x \text { and } z
\end{equation}

\noindent By integrating both sides of the second inequality in (\ref{obser conditions}) from $0$ to $T$, we obtain, 
\begin{equation}
V(x(T))-V(x(0)) \leq-\int_{0}^{T} z^{T} z d t
\end{equation}
Given that $V(x(T))\geq0$, we conclude that $V(x(0)) = x(0)^T P x(0)$ is an upper bound for the maximum output energy given the initial condition $x(0)$ \cite{boyd1994linear}. 
Assume  ($A$, $C$) are given.  Then, the second inequality can be formulated as the following LMI: 
$
P A + A^{T} P +C^{T} C \leq 0
$. Therefore, we get the best upper bound on the output energy by solving the following SDP in the variable $P$:
\begin{equation} \label{LMI observability}
\begin{aligned}
& \underset{P}{\text{minimize}}
& & x(0)^T P x(0) \\
& \text{subject to}
& & P>0 \\
&&& PA+A^T P + C^T C \leq 0
\end{aligned}
\end{equation}
This SDP can be solved analytically or using MATLAB CVX. The solution is exactly equal to the output energy $x(0)^T W_0 x(0)$, where $W_0$ is the observability  gramiam of the system: 
\begin{equation}
W_{\mathrm{o}} \triangleq \int_{0}^{\infty} e^{A^{T} t} C_{z}^{T} C_{z} e^{A t} d t
\end{equation}

Assuming $(A,C)$ are unknown,  the problem of finding the observability gramiam or the maximum output energy given the initial state $x(0) $ of a LTI system can be solved given only a single trajectory. The proposed method is based on finding a quadratic function that satisfies conditions in (\ref{obser conditions}) at every point along the given trajectory. Therefore, the conditions in (\ref{obser conditions}) can be represented numerically as follows: 
\begin{equation}\label{observability conditions tra}
\begin{gathered}
    x(i)^T P x(i) >0 \\ 
    2x(i)^T P \dot{x}(i)\leq - z(i)^T z(i)
\end{gathered}
\end{equation}

for $i=0,\dots,N$. Hence, instead of formulating the problem as an LMI problem as in (\ref{LMI observability}) given $(A,C)$, we can reformulate the problem as the following LP formulation provided that one single trajectory over a finite period of time is on hand:

\begin{equation} \label{LP observability}
\begin{aligned}
& \underset{p}{\text{minimize}}
& & (x(0)\oplus x(0))^T p \\
& \text{subject to}
& & L_1 p > 0 \\
&&& L_2 p  \leq -l_z
\end{aligned}
\end{equation}
where, $p = \text{vec}(P)$, $L_1$ and$L_2$ are defined in (\ref{L_1})and (\ref{L_2}), and $l_z$ is defined as: 
\begin{equation}\label{l_z}
l_z  = \begin{bmatrix}
z(0)^T z(0)\\
z(1)^T z(1)\\
\vdots \\ 
z(N)^T z(N)
\end{bmatrix}
\end{equation}

\noindent \underline{\textbf{Numerical example:}}
Given the following LTI system: $\dot{x} = Ax$ , $z = Cx$
\begin{equation} \label{LTI exmaple observability}
    A = \begin{bmatrix}
      0 & 1 \\ 
      -4 & -2
    \end{bmatrix} \quad \quad 
    C = \begin{bmatrix}
      0 &1 
    \end{bmatrix}
\end{equation}
A single trajectory at the initial state $x(0)=\begin{bmatrix} 2 &2 \end{bmatrix}^T$ over a finite time period $T =5 \text{sec}$ with a time step $dt =0.1$ is provided. The output $z$ and state vector $x$ at each time step can be measured. The derivatives of the state vector $\dot{x}$ can be calculated numerically. So, $N=50$. The LP problem of (\ref{LP observability}) is solved and the following $P$ matrix was the solution: 
\begin{equation}
\begin{bmatrix}
0.5000& 0.1250\\
0.1250&0.0625
\end{bmatrix}
\end{equation}

\noindent Which is exactly the same solution obtained by solving LMI (\ref{LMI observability}) where the full knowledge of the system dynamics is assumed. Therefore, from a single trajectory of the LTI system, the observability gramium matrix can be derived, then the maximum output energy of any given initial condition can be determined by $x(0)^T P x(0)$. 

\section{Data-driven Evaluation for bounds on output peak}\label{learn bounds on output peak}
The problem of deriving bounds on output energy $\norm{z(t)}$ can be formulated as a SDP \cite{boyd1994linear} when the system dynamics is known and the initial state $x(0)$ is given. Let $\mathcal{E}=\left\{\xi \mid \xi^{T} P \xi \leq 1\right\}$ be an invariant ellipsoid that contains $x(0)$ for the LTI system $\dot{x} =A x$ , $z= cx$. Then, 
\begin{equation} \label{bound on outpeak opt}
z(t)^{T} z(t) \leq \max _{\xi \in \mathcal{E}} \xi^{T} C^{T} C \xi
\end{equation}
In \cite{boyd1994linear}, the right hand side of (\ref{bound on outpeak opt}) can be expressed as the square root of the minimum $\delta$ subject to:
\begin{equation}\label{bound outpeak cons}
    \begin{bmatrix}
    P & C^T  \\
    C & \delta I 
    \end{bmatrix} \geq 0
\end{equation}
Therefore, given the initial state $x(0)$, the minimum bound of the output peak is the square root of the optimal value of the following SDP where $P$ and $\delta$ are decision variables for the optimization problem.
\begin{equation} \label{bound outpeak LMI}
\begin{aligned}
& \underset{P,\delta}{\text{minimize}}
& & \delta \\
& \text{subject to}
& & P>0 , \quad PA+A^T P  \leq 0\\
&&& x(0)^T P x(0) \leq 1 , \quad (\ref{bound outpeak cons})
\end{aligned}
\end{equation}

To obtain an optimal solution for the SDP of (\ref{bound outpeak LMI}), the system dynamics has to be known a prior. However, this problem can be solved without knowing the system matrices ($A,C$) where only a single trajectory is assumed to be available. Using our approach in sections \ref{learning lyapunov } and \ref{learning observability}, this problem can be formulated as an LP problem given the initial state and $(x,\dot{x},z)$ at every point along the given system trajectory. First, by Schur complement, (\ref{bound outpeak cons}) can be rewritten as follows: 
\begin{equation}
P-C^T C/\delta \geq 0
\end{equation}
this condition should be satisfied along the given trajectory $x(i)^T Px(i)-x(i)C^T C x(i)/\delta \geq 0$. As a result, the constraints of (\ref{bound outpeak LMI}) will be represented as follows: 

\begin{equation}
    \begin{gathered}
    x(i)^T P x(i) >0 \\
    2x(i)^T P \dot{x}(i) \leq 0\\
    x(0)^T P x(0) \leq 1 \\
    x(i)^T P x(i) -z(i)^T z(i)/\delta \geq 0
    \end{gathered}
\end{equation}
    
The fourth constraint is nonlinear, so let $\lambda = 1/\delta$ and instead of minimizing  $\delta$, $\lambda$ is maximized. Therefore, instead of formulating the problem as a SDP (\ref{bound outpeak LMI}), it can be formulated as the following LP problem given data points along the system trajectory: 
\begin{equation} \label{bound outpeak LP}
\begin{aligned}
& \underset{p,\lambda}{\text{maximize}}
& & \lambda \\
& \text{subject to}
& & L_1 p>0 , \quad L_2 p  \leq 0\\
&&& (x(0) \oplus x(0))^T p \leq 1 \\
&&&L_1 p - \lambda l_z \geq 0
\end{aligned}
\end{equation}
where, $p = \text{vec}(P)$, $L_1,L_2,l_z$ are defined in (\ref{L_1}),(\ref{L_2}) and (\ref{l_z}) respectively.
Given the initial state $x(0)$ and a single trajectory, the LP problem of (\ref{bound outpeak LP}) can be solved to obtain the upper bound of the output peak $\sqrt{1/\lambda}$ and the invariant ellipsoid $x^T P x \leq 1$ at which the maximum output of any trajectory within the ellipsoid will not exceed that upper bound.  

\noindent \underline{\textbf{Numerical example:}}

\begin{figure}[h]
    \centering
    \includegraphics[width=8cm,height=5cm]{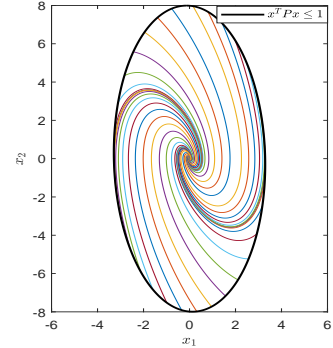}
    \caption{Data-driven invariant ellipsoid}
    \label{fig:2}
\end{figure}

 Using the same LTI system $\dot{x} =Ax,  z = Cx $  such that $A$ and $C$ are defined in (\ref{LTI exmaple observability}). Given the initial state $x(0)=[3,3]$ and a trajectory of the LTI system starting from $t=0 ~\text{sec}$ to  $T=5~\text{sec}$ with a time step $dt = 0.1~\text{sec}$. Then the LP problem (\ref{bound outpeak LP}) is solved to obtain the upper bound on the output peak and its corresponding invariant ellipsoid. The upper bound found to be ($\sqrt{1/\lambda}=3.2901$) and the invariant ellipsoid is $x^T P x \leq 1$ such that 
\begin{equation}
   P=  \begin{bmatrix}
    0.092453& 0.001486\\ 
   0.001486 & 0.015684 \\
    \end{bmatrix}
\end{equation}
Fig. \ref{fig:2} shows the invariant ellipsoid $x^T Px \leq 1$ that contains the initial state $x(0)$. As it can be seen from the figure, any trajectory starts at the boundary or within the defined ellipsoid remains inside it.

\begin{figure}[h]
    \centering
    \includegraphics[scale=0.5]{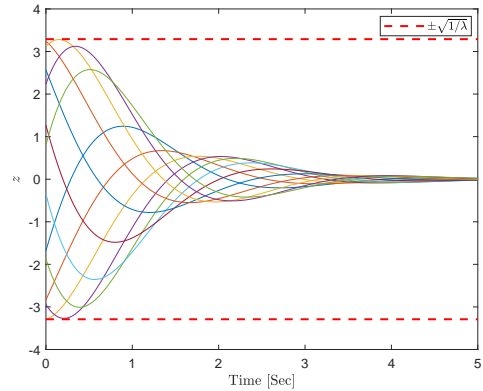}
    \caption{Data-driven upper bound for output peak}
    \label{fig:3}
\end{figure}

\noindent Fig. \ref{fig:3} shows the upper bound of the output peak and set of randomly selected trajectories starting at the boundary of the invariant ellipsoid. The figure shows that the output  $\norm{z(t)}$  of all trajectories inside the invariant ellipsoid remains below the obtained upper bound. 

\noindent The data-driven results of our proposed approach are almost identical to the solution of the SDP (\ref{bound outpeak LMI}) assuming the system dynamics is known. The upper bound for output peak derived from the SDP of (\ref{bound outpeak LMI}) is $3.2915$ and the invariant ellipsoid is $x^T P x \leq 1$ such that: 
\begin{equation} \label{P upper bound LMI}
P_{LMI}=\begin{bmatrix}
 0.092426& 0.001406 \\
 0.001406& 0.015873
\end{bmatrix}
\end{equation}
\section{Data-driven Evaluation for $\mathbf{L}_2$ gain and RMS gain}\label{learn L_2 norm}

For LTI system(\ref{LTI system}), $\text{L}_2$ gain is defined as: 
\begin{equation}
    \sup_{\norm{u}_{2} \neq 0} \frac{\norm{z}_{2}}{\norm{u}_{2}}
\end{equation} 
where the supremum is over all nonzero trajectories of the LTI system starting from $x(0)=0$. In \cite{boyd1994linear}, If there exists a quadratic function $\mathcal{E}=\zeta^T P \zeta$, $P>0$ and for all $t\geq0$,
 \begin{equation}\label{rbl equation}
 \frac{d}{\text{dt}} V(x)+z^{T} z-\gamma^{2} u^{T} u \leq 0 \quad \forall x \text { and } u \text { satisfying (\ref{LTI system}) }
 \end{equation}
such that $\gamma\geq0$, then the $\mathbf{L}_2$ gain of the LTI is less than $\gamma$. To show that, integrate both sides of (\ref{rbl equation}) from $0$ to $T$ with $x(0)=0$, we obtain
\begin{equation}\label{rbl equation integral}
    V(x(T))+\int_{0}^{T}\left(z^{T} z-\gamma^{2} u^{T} u\right) dt\leq 0
\end{equation}
$V(x(T))\geq 0 $,  so we conclude that $\int_{0}^T z^T z dt\leq \gamma^2 \int_0^T u^T u dt  $. 
\begin{itemize}
    \item Take the limit $T\to \infty$, then the system's $\mathbf{L}_2$ gain is less than $\gamma$
    \item Divide by $T$ and take the limit $T \to \infty$, then $\frac{\mathbf{RMS}(z)}{\mathbf{RMS}(u)} \leq \gamma$, Therefore, the system's $\mathbf{RMS}$ gain is less than $\gamma$
\end{itemize}

\noindent Given the system dynamics ($A,B,C$), equation (\ref{rbl equation}) can be written as follows:
\begin{equation}\label{rbl equation expansion}
x^T(PA+A^T P + C^T C)x+2x^T PB u -\gamma^2 u^T u \leq 0 
\end{equation}
that can be formulated as the following LMI: 
\begin{equation}\label{rbl LMI}
\begin{bmatrix}
    PA+A^T P + C^T C & PB \\ 
    B^T P   & -\gamma^2 I 
\end{bmatrix} \leq 0 
\end{equation}
Hence, the smallest upper bound for the LTI system's $\mathbf{L}_2$ or $\mathbf{RMS}$ gain can be obtained by minimizing $\gamma$ over the variables $P$ and $\gamma$ 
while satisfying (\ref{rbl LMI}) and $P>0$. This method is based on the full knowledge of the system's dynamics ($A,B,C$). Our approach can be also used here to solve this problem given only a single trajectory of the LTI system starting from $x(0)=0$. Instead of solving the SDP problem of (\ref{rbl LMI}), the problem can be reformulated as an LP problem by minimizing $\beta = \gamma^2$ while the conditions $P>0$ and (\ref{rbl equation}) at any data point $i$ along the given trajectory are satisfied in the following format:
\begin{equation}
\begin{gathered}
        x(i)^T P x(i) > 0 \\
        2x(i)^T P \dot{x}(i) + z(i)^T z(i)- \beta u(i)^T u(i) \leq 0
\end{gathered}
\end{equation}
where $\gamma^2$ is replaced by $\beta$ in both the objective and the second condition to preserve the linearity. Therefore, the Linear program can be written as: 

\begin{equation} \label{rbl LP}
\begin{aligned}
& \underset{p,\beta}{\text{minimize}}
& & \beta \\
& \text{subject to}
& & L_1 p>0 \\
&&& L_2 p +l_z -\beta l_u \leq 0 \\
\end{aligned}
\end{equation}
where $p=\text{Vec}(P)$ and $L_1$,$L_2$ and $l_z$ are defined by (\ref{L_1}),(\ref{L_2}) and (\ref{l_z}) respectively. $l_u$ is: 
\begin{equation}\label{l_u}
    l_u  = \begin{bmatrix}
u(0)^T u(0)\\
u(1)^T u(1)\\
\vdots \\ 
u(N)^T u(N)
\end{bmatrix}
\end{equation}

\noindent \underline{\textbf{Numerical example:}}
 Given the following LTI system: 
\begin{equation}\label{LTI system rbl}
\begin{aligned}
\dot{x} &= \begin{bmatrix} 0 & 1 \\ 
-1 &-2
\end{bmatrix}x + \begin{bmatrix}
    1 \\2 
\end{bmatrix}u \\ 
z &= \begin{bmatrix}
    4 & 1
\end{bmatrix} x
\end{aligned}
\end{equation}

Assuming that ($A,B,C$) are known. So,we can get the the system $\mathbf{L}_2$ gain by solving the following SDP: \begin{equation} 
\begin{aligned}
& \underset{P,\gamma}{\text{minimize}}
& & \gamma \\
& \text{subject to}
& & P>0, \quad (\ref{rbl LMI})
\end{aligned}
\end{equation}
So, $\mathbf{L}_2$ gain of the system is $\mathbf{15}$. Using our approach, $\gamma$ can be obtained by exciting the system with a unit step input from $0$ to a final time $T$, then measure the input-output ($u$,$z$) and states ($x,\dot{x}$) data of a single trajectory. Therefore,  the linear program (\ref{rbl LP}) can be solved to get $\gamma=\sqrt{\beta}$. The time step used is $0.01$ sec. Table \ref{table:1} shows that as the length of the trajectory increases, the result's accuracy will be improved.
\begin{table}[h]
  \caption{Learned $\mathbf{L}_2$ gain}
    \centering
    \begin{tabular}{|c|c|}
    \hline
       Final time $T$ [Seconds] & Learned $\mathbf{L}_2$ gain ($\gamma$) [SDP solution=15]\\
    \hline \hline
    2 &  7.36147\\
    \hline 
      4 &  12.17092\\
    \hline 
     6 &  14.30461\\
    \hline 
     8 &  14.86621\\
    \hline 
     10 &  14.97684\\
    \hline 
     12 &  14.99619\\
    \hline 
    14 &  14.9994\\
    \hline 
    16 & 14.99990\\
    \hline
    \end{tabular}
    \label{table:1}
\end{table}

Note that $\mathbf{L}_2$ gain of the LTI system is the $\mathbf{H}_{\infty}$ norm of its transfer function $C(sI-A)^{-1}B$. Therefore, using our approach, the $\mathbf{H}_{\infty}$ norm of the LTI system can be learned accurately using only a single trajectory and without need to know its dynamic matrices.

\section{Conclusion and future work}
In this work, a number of data-driven techniques that aim to evaluate various metrics of LTI systems such as the upper bound on the output 's energy and peak as well as the $\mathbf{L}_2$ and $\mathbf{RMS}$ gains were introduced. In addition, the exact and approximate construction of the Lyapunov function of LTI systems were proposed. To demonstrate the proposed methodologies, a number of numerical examples were given and thoroughly discussed. As for future work, we are extending the proposed data-driven construction for the Lyapunov function to a non-quadratic Lyapunov function for hybrid linear time-invariant systems and specific classes of nonlinear systems.  Finally, we are extending the proposed data-driven methodologies to linear systems where uncertainties happen and the measurements are noisy.

\addtolength{\textheight}{-12cm}   

\bibliographystyle{ieeetr}

\bibliography{main.bib}

\end{document}